\documentclass[aps,prd,prabib,twocolumn,showpacs,nofootinbib]{revtex4}
\usepackage{graphicx} \usepackage{amsmath} \usepackage{amssymb}
\usepackage{amsfonts} \usepackage{bm}
\usepackage{dcolumn}

\begin{document}

\newcommand{\be}{\begin{equation}} \newcommand{\ee}{\end{equation}}
\newcommand{\bea}{\begin{eqnarray}}\newcommand{\eea}{\end{eqnarray}}

\title{Hidden scale  in quantum mechanics}

\author{Pulak Ranjan Giri} \email{pulakranjan.giri@saha.ac.in}

\affiliation{Theory Division, Saha Institute of Nuclear Physics,
1/AF Bidhannagar, Calcutta 700064, India}

\begin{abstract}
We show that the intriguing localization of a free particle
wave-packet is possible  due to a hidden scale present in the
system. Self-adjoint extensions (SAE) is responsible for introducing
this scale in quantum mechanical models through the nontrivial
boundary conditions. We discuss a couple of  classically scale
invariant  free particle systems to illustrate the issue. In this
context it has been shown that  a free quantum particle moving on a
full line may have localized wave-packet around the origin. As a
generalization, it has also been shown that particles  moving on a
portion of a plane or on a portion of a three dimensional space can
have unusual localized wave-packet.

\end{abstract}



\date{\today}

\maketitle
\section{Introduction}
In quantum mechanics usually  the bound state system described by a
Hamiltonian must have a scale in the Hamiltonian in order to
localize it in a region of space. This is the reason a particle with
only kinetic term is a free particle with wave-function spreading
throughout the space with equal probability. Even a particle with a
potential \cite{giri1,giri2,giri3,giri4}, which transforms the same
way the kinetic part transforms under scale transformation
$\boldsymbol{r}\to\alpha\boldsymbol{r}$, $t\to\alpha^2t$, does not
usually possess any bound state \cite{landau}.

Despite this scale invariance in some problems \cite{camblong} one
can still expects bound state solution when quantization of the
classical system is preformed. Because the process of quantization
may introduce a scale into the system. SAE \cite{reed} is one way of
introducing a scale in the system, thus leading to a quantum
mechanical anomaly \cite{giri1,giri2,giri3,giri4,camblong}. Thus
although we don't see the scale in the Hamiltonian, it is actually
hidden in the boundary condition. SAE has been a rigorous method to
find the most general boundary conditions for a quantum mechanical
model so that the operator, for example the Hamiltonian, becomes
self-adjoint. For the Hamiltonian it is necessary to be
self-adjoint, because otherwise the time evolution of the quantum
states generated by $\mathcal{U}=\exp\left(-iHt\right)$ \cite{reed}
will not be unitary. Unitarity  is essential to keep the norm of the
states unchanged through out the transformation. The other
importance of the self-adjointness is that the eigen-values are
guaranteed to be real.

Hidden scale problem, quantum anomaly and the implications of
self-adjoint extensions, all these three can be found  in the case
of a free particle dynamics. Note that we call a particle free in
the sense that the potential for the particle $V=0$, i.e., it has
only kinetic part in the Hamiltonian, $H=\boldsymbol{p}^2/2M$.
Although the form of the Hamiltonian is the simplest of all, it
raises lot of intriguing facts when viewed as a Hamiltonian of a
localized wave packet. For example, the localization of a free
particle on a half line \cite{bonneau,gar} is such an interesting
problem, where SAE gives rise to bound state solutions by
introducing a length scale into the system. Similarly for a particle
confined on a whole plane can have bound state solution, once
inequivalent quantization is made \cite{cirone,kow}. The largest
possible space dimensions in which a free particle can have bound
state due to inequivalent quantization is $N=3$. Beyond three
dimensions the quantum centripetal inverse square potential arising
from pure kinetic term does not allow  the localization of the
wave-packet.

In this letter, in Sec. II we will discuss  the problem of binding a
free particle on a whole line by generalizing the problem of a
particle on a half line.
In Sec. III we discuss about a particle moving on a portion of a
plane (see FIG. 2) and also discuss the problem of a free particle
on a plane (see FIG. 1) in the context of hidden scale problem.
Finally a free particle moving in some region of a three dimensional
space has been shown to possess a bound state in Sec. IV. All these
three problems are scale invariant due to the absence of any
potential in the Hamiltonian. However the fact that very unusual
bound state does exists in all these three cases was not known in
the literature as far as our knowledge is concerned. We conclude in
Sec. V.
\begin{figure}
\includegraphics[width=0.45\textwidth, height=0.15\textheight]
{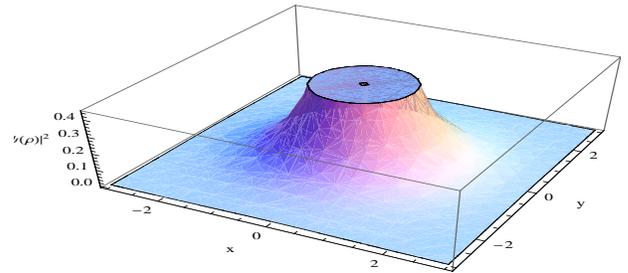}
\caption {(color online) Bound state probability density $|\psi(\rho)|^2$ for a
  particle on a plane  with length scale $L=1$ has been plotted
  as a function of $x, y$. The probability density is zero at the origin
($0,0$), which is indicated as the black spot on the top of the figure.}
\end{figure}
\section{Particle on a full line}
Before discussing  the problem of a free particle moving on a full
line let us first review  the problem on a half line
\cite{bonneau,gar}. Because then particle on a full line is just a
generalization. It is known that the free particle on a half line
can be made self-adjoint and there exist a bound state of the
particle. The 1-dimensional Hamiltonian for the particle in the
interval ${\mbox x}\in[0,\infty)$ is of the simple form
($\hbar^2=2M=1$)
\begin{equation}
H_{1D}= -\partial^2_{\mbox x}\,.\label{1H1}
\end{equation}
We are interested in the bound state problem for the particle. The
Hamiltonian is manifestly scale covariant under the transformation
${\mbox  x}\to \alpha {\mbox x}$, $t\to\alpha^2t$. So there is no
scale in the problem and it suggests that the particle does not have
any bound state \cite{camblong}. But the inequivalent quantization
of the system with the self-adjoint domain
\begin{equation}
\mathcal{D}^L_{1D}=\{\psi(\mbox x)\in\mathcal{L}^2(d\mbox
x),\psi'(0)= L^{-1}\psi(0)\}\,,
 \label{1H1b}
\end{equation}
allows us to get a bound state solution with energy  eigenvalue and
eigenfunction respectively given by
\begin{equation}
E_{1D}= - L^{-2},~~ \psi(\mbox x)_{1D}= \sqrt{2L^{-1}}\exp(-L^{-1}\mbox x)\,.
 \label{1H1s}
\end{equation}
where $L$ has to be  positive and finite in order to make the
solution $\psi(\mbox{x})$ square-integrable. The hidden scale $L$, called
the self-adjoint extension parameter,  breaks the scale invariance
of the system. This is a simple quantum mechanical example of
scaling anomaly. The probability density for the wave-packet confined on a
half line has been shown in FIG. 3.
\begin{figure}
\includegraphics[width=0.50\textwidth, height=0.15\textheight]
{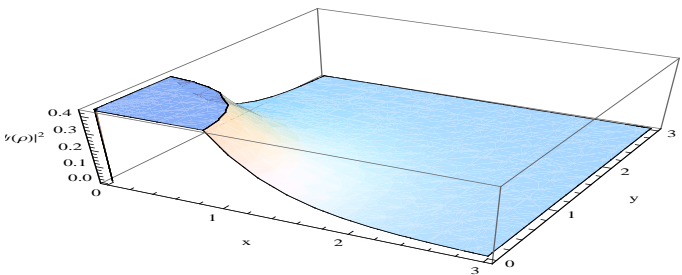}
\caption {(color online) Bound state probability density $|\psi(\rho)|^2$ for a
  particle on a part of the plane (first quadrant)
with length scale $L=1$ has been plotted
  as a function of $x, y$. The probability density is zero at the origin
($0,0$).}
\end{figure}

We now generalize the same problem by considering it on a full line,
$\mbox{x}\in(-\infty,\infty)$ instead on a half line.
The Hamiltonian  now possesses reflection symmetry in addition to
its scale invariance contrary to the half line case, which had only
scale invariance. We can exploit the the reflection symmetry of the
problem to reduce it on the form (\ref{1H1})
by using  the  transformation  $\mbox{z}= |\mbox{x}|$.  So the
analysis will be same, but the normalization constant of the bound
state wave-function will now change due to reflection symmetry in
the problem. The bound state solutions are
\begin{equation}
E_{1D}= - L^{-2},~~ \psi(|\mbox{x}|)_{1D}=
\sqrt{L^{-1}}\exp(-L^{-1}|\mbox{x}|)\,.
 \label{1H2s}
\end{equation}
Note the simplicity of the result (\ref{1H2s}), but despite its
simplicity it has remained unnoticed so far. It is however know for
a long time that particle on a line with $\delta$-function potential
has bound state solution \cite{cirone,note1}. In fact the result is
same as what we have obtained without  any potential but using SAE.
The probability density has a pick at the origin, which has been
shown in FIG. 5.
\begin{figure}
\includegraphics[width=0.45\textwidth, height=0.15\textheight]
{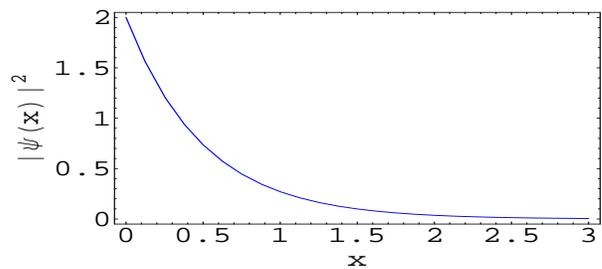} \caption {(color online) Bound state probability
density $|\psi(\mbox{x})|^2$ for a
  particle on a half line ($\mbox{x}\in[0,\infty)$) with length scale $L=1$ has been
  plotted as a function of $\mbox{x}$. The probability density is maximum at the
  origin.}
\end{figure}
\section{Particle  on a plane}
To show the ring shaped localization of a free particle
wave-function around the origin of a plane \cite{cirone,kow,birula}
due to the hidden scale, we consider a particle of mass $M$ on
$x$-$y$ plane. The Hamiltonian of the system can be written in term
of a 2-dimensional Laplacian $H_{2D}= -\boldsymbol{\nabla}^2$. In
polar co-ordinates ($\rho,\phi$) the radial eigenvalue equation with
eigen-value $E$ can easily be separated with the radial Hamiltonian
\begin{equation}
H^\rho_{2D}= -\partial^2_\rho - 1/\rho\partial_\rho +
m^2/\rho^2\,,\label{2H1_r}
\end{equation}
where $\partial_\rho$  and $\partial^2_\rho$ is the 1-st and 2-nd
order derivative w.r.t $\rho$ respectively and $m=0,\pm1,\pm2,...$
is the angular momentum quantum number. Usual practice is to define
a very restricted symmetric domain for this system so that it can be
extended to a self-adjoint domain. One of the possible  domains over
which the Hamiltonian is symmetric is of the form
\begin{equation}
\mathcal{D}_{2D}=\{\psi(\rho)\in\mathcal{L}^2(\rho
d\rho),\psi(0)=\psi'(0)=0\}\,. \label{domain1}
\end{equation}
The domain $\mathcal{D}_{2D}$ is so restricted that it fails to make
$H^\rho_{2D}$ self-adjoint. Then one seeks for a SAE. Using von
Neumann's method it can be shown that the domain over which the
Hamiltonian $H^\rho_{2D}$ is self-adjoint is of the following form
\begin{eqnarray}
\nonumber
\mathcal{D}_{2D}^\Sigma(L^{-2})=~~~~~~~~~~~~~~~~~~~~~~~~~~~~~~~~~~~~~
\\\{\mathcal{D}_{2D} +
\psi(\rho,L^{-2},\Sigma)|\psi(\rho,L^{-2},\Sigma)\in
\mathcal{D}_A\}\,, \label{2Hds}
\end{eqnarray}
where $\mathcal{D}_A$ is the domain of the operator
${H^\rho_{2D}}^*$, which is  adjoint to $H^\rho_{2D}$. The
dimensionless parameter $\Sigma\in\mathbb{R}(\mbox{mod}~2\pi)$ is
called the SAE parameter. Note that the dimension-full constant
$L\in\mathbb{R}^+$ is incorporated into the domain
$\mathcal{D}_{2D}^\Sigma(L^{-2})$ through the elements
$\psi(r,L^{-2},\Sigma)$ of the deficiency space, which is spanned by
the solutions of the equation
\begin{eqnarray}
({H^\rho_{2D}}^* \mp iL^{-2})\psi_\pm(\rho,L^{-2})=0\,.
\end{eqnarray}
The element $\psi(\rho,L^{-2},\Sigma)$ is explicitly written as
$\psi(\rho,L^{-2},\Sigma)= \psi_+(\rho,L^{-2})+
\exp(i\Sigma)\psi_-(\rho,L^{-2})$. Now the system defined by
$H_{2D}$ and  $D_{2D}^\Sigma$ has a length scale $L$, hidden in the
boundary condition. The bound state for the system is now exists for
$m=0$ wave and it will now depend on two independent parameters
$\Sigma$ and $L$. The bound state energy $E(L^{-2},\Sigma)$ has
certain interesting features, for example it is periodic in
$\Sigma$,
\begin{eqnarray}
E(L^{-2},\Sigma)= E(L^{-2},\Sigma+ 2\pi)\,.
\end{eqnarray}
So the bound state energy $E(L^{-2},\Sigma)$ can be written in terms
of a periodic function $\mathcal{F}(\Sigma)\in\mathbb{R}^+$ as
\begin{eqnarray}
E(L^{-2},\Sigma)= -L^{-2}\mathcal{F}(\Sigma)\,. \label{2b}
\end{eqnarray}
The exact form of the function $\mathcal{F}(\Sigma)$ can be found
from the domain $\mathcal{D}_{2D}^\Sigma(L^{-2})$. The bound state
eigenfunction for $\mathcal{F}(\Sigma)=1$ is of the form
\cite{giri1}
\begin{eqnarray}
\psi(\rho)= \frac{1}{\sqrt{\pi}}L^{-1}K_0(L^{-1}\rho)\,,\label{2bs}
\end{eqnarray}
where $K_0$ is the modified Bessel function \cite{abr}, which has
logarithmic  divergence at the origin but the probability density
obtained from it  goes to zero at origin, which has been shown in
FIG. 1.
\begin{figure}
\includegraphics[width=0.40\textwidth, height=0.35\textheight]
{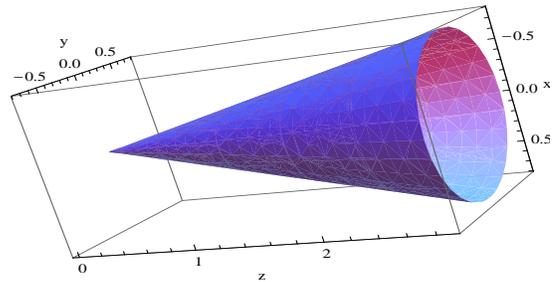} \caption {(color online) A  particle with $V=0$
is confined in
 a part of the 3-dimensional space specified by $r\in[0,3], \theta\in[0,
  \pi/12], \pi\in[0,2\pi]$. The probability distribution $|\psi(r)|^2$
of the particle as a function of the radial co-ordinate will look
like FIG. 3, where probability density for a particle moving on a
half line has been plotted. The reason for this similarity is
obvious from the fact that for $\theta=0$, the solid angle reduces
to a half line.}
\end{figure}

We now consider the situation, where the particle is moving on a
part of the plane not on a whole plane and ask the question whether
the method of SAE is still capable of binding the particle on the
restricted region of the plane, for example in the region specified
by $\rho\in[0,\infty),\phi\in [0, 2\pi\beta]$, where $0\leq\beta\leq
1$. This problem can be easily solved once the eigen-value  equation
for the angular operator
$\boldsymbol{L}^2=-\partial^2/\partial\phi^2$ is solved. But we
don't need to explicitly solve the angular part for our discussion.
What we need to know is that whether there exists any eigenvalue
within the interval $0\leq{\tilde
 m}^2< 1$, because then only we can expect bound state solutions. One can
easily convince oneself that $\chi(\phi)= \frac{1}{2\pi\beta}$ is
one of the eigen-functions  of the operator $\boldsymbol{L}^2$ with
eigen-value $\tilde m= 0$. Note that $\chi(\phi)$ has the time
reversal symmetry \cite{kow}. Thus  the radial Hamiltonian for
$\tilde m=0$ wave will be $H_{2D}^\rho= -\partial^2_\rho -
1/\rho\partial_\rho$, which has been shown in  (\ref{2b}) and
(\ref{2bs})  to possess bound state solution. The probability
density for the radial eigen-function has been plotted in FIG. 2.
for $\beta=1/4$.

\begin{figure}
\includegraphics[width=0.45\textwidth, height=0.15\textheight]
{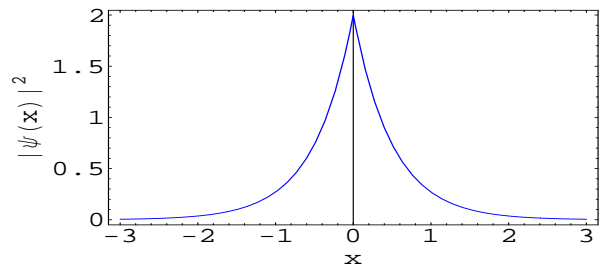} \caption {(color online) Bound state probability
density $|\psi(\mbox{x})|^2$ of a
  particle on a full line ($\mbox{x}\in(-\infty,\infty)$)  with length scale $L=1$
  has been plotted as a function of $\mbox{x}$. The probability density is maximum at
  the origin.}
\end{figure}
\section{Particle in N-dimensions}
We consider a free particle  moving in $N$ dimensional flat space.
The Hamiltonian for the system is  then written in the following
form
\begin{equation}
H_{ND}= -\boldsymbol{\nabla}^2\,. \label{ND1}
\end{equation}
Since (\ref{ND1}) has only kinetic term, it gives classically scale
invariant action under the scale transformation ${\bf r}\to
\varrho\bf r$, $t\to \varrho^2 t$. Thus, usually it does not have
any bound state solutions and only has free particle solutions
$\psi(\boldsymbol{r})=\exp(\pm i\boldsymbol{k}.\boldsymbol{r})$,
where $\boldsymbol{k}$ is the wave vector of the particle. The
energy for the free particle eigen-function, $E= \boldsymbol{k}^2$,
is continuous. We now seek for a nontrivial  solution of the
Schr\"{o}dinger eigenvalue equation for the Hamiltonian $H_{ND}$. In
spherical polar co-ordinates ($r,\phi_i$) the radial Hamiltonian can
be separated in the following form
\begin{eqnarray}
H^r_{ND}=
-\frac{1}{r^{N-1}}\frac{d}{dr}\left(r^{N-1}\frac{d}{dr}\right)+
\frac{l(l+N-2)}{r^2}
\end{eqnarray}
We can now use the transformation $R(r)= r^{-(N-1)/2}\chi(r)$ on the
Schr\"{o}dinger eigenvalue equation  $H^r_{ND}R(r)= E_{ND}R(r)$. The
Hamiltonian of the transformed eigenvalue equation
$\mathcal{H}^r_{ND}\chi(r)= E_{ND}\chi(r)$ has the very familiar
form $\mathcal{H}^r_{ND}= -\partial^2_r+ \mbox{g}/r^2$, with
$\mbox{g}= l(l+N-2) +3- N(N-4)$. It can be shown that
$\mathcal{H}^r_{ND}$  have only one bound state for
$-1/4\leq\mbox{g}<3/4$ \cite{giri1}. One can check that for $l=0$
and $2\leq N <4$ the effective coupling constant $\mbox{g}$ lies in
the specified interval.  Thus only s-waves for $N=2$ and $3$ support
bound state \cite{cirone}. The bound state solutions can be found
from the self-adjoint  domain
\begin{eqnarray}
\nonumber
\mathcal{D}_{ND}^\Sigma(L^{-2})=~~~~~~~~~~~~~~~~~~~~~~~~~~~~~~~~~~~~~~
\\ \{\mathcal{D}_{ND} +
\psi(r,L^{-2},\Sigma)|\psi(r,L^{-2},\Sigma)\in \mathcal{D}_{NA}\}\,,
\label{NHds}
\end{eqnarray}
where
$\mathcal{D}_{ND}=\{\psi(r)\in\mathcal{L}^2(r^{N-1}
dr),\psi(0)=\psi'(0)=0\}$
and $\mathcal{D}_{NA}$ is the domain of the adjoint Hamiltonian
${H^r_{ND}}^*$. Note that the scale $L$ is within the domain
$\mathcal{D}_{ND}^\Sigma(L^{-2})$, which has been introduced at the
time of SAE. The bound state solution  will now depend on the value
of $\mbox{g}$ in the interval.

The bound state problem on a plane  ($N=2$) has been discussed in the
previous section. Therefore  we
now concentrate the three dimensional ($N=3$) problem. The Hamiltonian
 simply becomes $\mathcal{H}_{3D}^r= -\partial^2_r$, because the dimensionless
 coupling $\mbox{g}=0$, for $l=0$ and $N=3$. It is now a one dimensional
problem on a half line, which has been discussed in Sec. II. The probability
density for the wave-packet will be like FIG. 3.

One can also consider the situation where a particle is moving only
in a portion of a 3-dimensional space, for example in  the region
$r\in[0,\infty)$, $\theta\in[0,\gamma\pi)$,  $\phi[0,2\pi]$, where
$0\leq \gamma\leq 1$. To solve this problem we need to solve the
angular part. In fact in our purpose it is enough to know the
coupling constant of the inverse square centrifugal term. One can
convince oneself that $Y(\theta,\phi)= C$ (complex valued constant)
is the trivial eige-function of  $\boldsymbol{L}^2$ with eigenvalue
$0$. Once again it reduces to a problem on a half line, discussed in
Sec. II. In FIG. 4 particle confinement in a solid angle has been
considered, where the probability density looks like FIG. 3.

\section{Conclusion}
Free particle Hamiltonian usually does not possess any bound state
solution due to the absence of any scale in the problem.  But we
have discussed that the scale, hidden in the boundary condition, may
be responsible to localize the wave packet. As an example we have
discussed the known problem of particle on a half line and particle
on a plane to show that the scale hidden within the boundary
condition is responsible for localizing the wave-packet. We have
also discussed that the free particle on a full line does have bound
state if inequivalent quantization is considered. It is however
known that a $\delta$-function potential can bind a particle on a
full line. So one may think that the SAE induces a $\delta$-function
potential in the system. Similar confinement of the wave packet has
been shown to hold for the case of a particle moving on a portion of
a plane and in a portion of a 3-dimensional space. These types of
very unusual localized wave-packet in some portions of a two and
three dimensional spaces does not seem to have appeared in
literature.
\section{Acknowledgment}
We are grateful to P. B. Pal for some useful discussions and
suggestions regarding Sec. II.

\end{document}